# A Possible Solution to the Cosmological Constant Problem By Discrete Space-time Hypothesis

H.M.Mok

Radiation Health Unit, 3/F., Saiwanho Health Centre,

Hong Kong SAR Govt, 28 Tai Hong St.,

Saiwanho, Hong Kong, China.

**Abstract** The cosmological constant problem is explained by a theory based on the discrete space-time hypothesis. The calculated $\lambda$ value is of the order of $10^{-52}[m]^{-2}$ or equivalent to about $\Omega_\lambda = 0.7$. It is in excellent agreement with the Type Ia SN observational data and recent results of BOOMERANG and MAXIMA. Our theory also implies that the quantization of the space-time metric $g_{mn}$ is not necessary since it is not a fundamental field. The divergence problem of quantum gravity is then of no interest. Cosmic inflation is given out as a consequence of the theory and the universe is found to be alternatively dominated by the cosmological constant and the mass density at different cosmic time period. Our calculation also shows that $\rho_m$ is of similar order of magnitude as $\rho_v$ in the present universe but it is just a coincidence. This result supports the anthropic principle.

The international collaboration on the High-Z SN Ia observation, which was aimed at measuring the cosmic deceleration and global curvature, found that the universe is accelerating instead of decelerating. The observational results from both teams, Perlmutter [1] and Schmidt [2], indicated that there is a non-vanishing cosmological constant ($\lambda$) in our universe. The value of $\Omega_\Lambda (\lambda c^2 / 3 H_0^2)$ is a few tenths of the critical mass density and, when compared with the Planck scale or electroweak scale, is many order of magnitudes smaller than that expected in quantum field theory. Recent observations by the BOOMERANG and MAXIMA also support such finding [3]. This cosmological constant problem is one of the mysteries of both cosmology and particle field theory. Viable approach, such as quintessence, anthropic principle and higher dimensional brane world solution are under investigation (Recent concise comments on such approaches can be found on [4]). However, up to now, there is still no satisfactory prediction on the $\lambda$ value using such attempts. On the other hand, the theoretical estimate of the $\lambda$ value by the quantum field theory is based on the spontaneous symmetry breaking process provided by the Higgs mechanism [5]. Although such theory is successful in most of the particle experiments, it is an *ad hoc* mechanism and lacks detail understanding. For example, the reason of existence of a scalar field in the vacuum is still not known. The present situation on the $\lambda$ problem urges us to consider both problems together in a new direction.

Although it is strange that there exists so large discrepancy between the theoretical and experimental $\lambda$ value, we may not be too unfamiliar with such situation. If we compare the $\lambda$ problem with the property of mass density, similar characteristic can be found. Due to the atomic structure of matter, mass density is relatively small in macroscopic scale (around $10^3 kg/m^3$ for common materials) but extremely large inside the atomic nucleus (around $10^{18} kg/m^3$). The case for the cosmological constant, also, is very small in macroscopic scale but extremely large when quantum field consideration (i.e. microscopic scale) has been put in. If the $\lambda$ problem is analogous to the case of matter density, it indicates that some kind of discrete structure may exist in the vacuum. Since the vacuum and space-time are indistinguishable, such discrete vacuum properties may further imply that the space-time itself is also discrete in nature. Besides, there is no evidence that space-time is smooth and continuous in extreme microscopic scale. In quantum gravity, space-time is expected to have a very different geometrical structure in Planck scale, such as the existence of wormholes and space-time foams. Also, the concept of discrete space-time is not new. T.D.Lee [6], G.t'Hooft [7] and others [8] had considered such possibility in resolving the UV/IR divergence problem in quantum gravity but they had not related it to the problem of the cosmological constant at that time.

If space-time is really discrete in nature, the cosmological constant described by the quantum field theory can be just the situation inside the basic constituents of space-time and is many order of magnitudes larger than the macroscopic observational data. The quantum field theory description and the cosmological observations of the vacuum energy density may then both be correct on its corresponding length scales. Furthermore, the spontaneous symmetry breaking of the scalar field may be corresponding to the phase transition of such space-time "condensate" like structure.

Let us show the above idea in a paradigm. If we imagine our 3-D space as a 2-D elastic membrane, which is commonly use for the illustration of cosmic expansion, the discrete space-time is corresponding to a membrane which is not smooth but has its atomic structure (in fact, a physical membrane is made of atoms). The gravitational field for the 2-D creatures living on the membrane is the properties of deforming the membrane by mass and general relativity is then a kind of continuous theory of elasticity. They may find that the space is smooth in macroscopic scale but has its microscopic structure. Also, the "internal energy" of the membrane (It may be viewed as the vacuum energy by the 2-D creatures.) is extremely large in microscopic scale (i.e. the nuclear energy or the atomic bonding energy) but small in macroscopic scale and such "internal energy" of his space does not curve the space-time structure. This simple model gives the properties of our cosmological constant.

Based on the above arguments, we postulate that : (1) Space-time is discrete in nature and its fundamental unit is of the order of Planck scale; (2) The space-time forms a kind of phase (or say "condensate") with its constituents; (3) The scalar field plays both the role as the order parameter of such space-time phase and the wavefunction of its constituents as Cooper pairs in superconductivity [9,10] (We have to remark that such "condensate" may not be exactly the same as in usual understanding but has similar properties that is useful to draw analogy between them). Some important consequences can be directly followed from these postulates. Firstly, the Higgs particle will be just a kind of excited state of the individual space-time constituent. Secondly, the divergence problem of the quantum gravity is of no interest since the space-time metric $g_{mn}$ is a collective effect of the space-time constituents (like the strain tensor of elasticity of a membrane) and is not a fundamental field itself. General Relativity is also not a fundamental field theory but is just a collective description. Using the language of the paradigm described above, there is no need to insist in quantizing the wave propagation of the strain tensor on a 2-D membrane since the atomic lattice vibration is the one that needed to be quantized (phonons). Back to our case, it is the field of the space-time constituents (or say the scalar field) and its lattice like vibration energy that need to be quantized. Therefore, the problem of quantum gravity is reduced to the quantization of the vibration energy of the space-time constituents and in some sense we can say that the quantization of Higgs field is already part of a quantum gravity theory! It seems that the concept in condensed matter physics is useful in building the quantized model of the space-time structure.

If we take the form of the scalar field potential as

$$V = V_0 - m^2 f^* f + g(f^* f)^2 \qquad (1)$$

where $m^2 > 0, g > 0$ and assuming that $V_0 = 0$ (They has its usual meaning as in the electroweak theory. SM Higgs is assumed here for simplicity and also to avoid unnecessary complexity). The energy density at broken symmetry $r_h$ (which will be shown to be different from the macroscopic vacuum energy density in cosmological observation) of the scalar field is given by $r_h = -m_{EW}^4$ [5], where $m_{EW}$ is the electroweak mass scale. Since the shape of the potential around the minimum $f \neq 0$ determines the mass of the quantum particle and, in our case, it is equal to the Higgs mass, the energy density $r_h$ expression can then be interpreted as the contribution of the scalar field of mass $-m_{EW}$ with number density $m_{EW}^3$. The negative sign of $r_h$ makes it behaves as the microscopic vacuum energy density with value $m_{EW}^4$ (We use the sign convention that +ve vacuum energy density corresponding to –ve normal energy density.). Also, because the scalar field is the wavefunction of the space-time constituents, $r_h$ acts as a kind of

"binding energy" (or internal energy) of the constituents and, as mentioned above, such "binding energy" will not curve the space-time since it behaves as the internal energy of the space-time structure. If phase transition happens on the space-time "condensate", this "binding energy" can be released as positive energy and is important in early universe evolution that will be discussed later.

Due to the energy density $r_h$, pressure will be created on each space-time constituents and with the value equals to

$$P = r_h = -m_{EW}^4 \qquad (2)$$

The sign on the RHS of equation (2) shows that the pressure is negative (i.e. behave as a stretching force on individual space-time constituent but as a "binding force" between the space-time constituents.). This value is what we expect from quantum field theory [5]. However, as discussed above, the space-time is postulated to be discrete and its fundamental unit is of the order of Planck scale with estimated density of about $m_{EW}^3$. Therefore, the macroscopic vacuum energy density should be weighed by a factor of $(m_{EW}/M)^3$ to average it out on an ideally smooth macroscopic space scale (just as average out the nuclear mass density to get the macroscopic mass density of matter by considering the atomic spacing), where $M$ is the Planck mass. The macroscopic vacuum energy density $r_v$ will become

$$r_v = \left(\frac{m_{EW}}{M}\right)^3 m_{EW}^4 = \left(\frac{m_{EW}^7}{M^3}\right) \qquad (3)$$

The cosmological constant is then equal to

$$\lambda = k r_v \sim \left(\frac{m_{EW}^7}{M^5}\right) \qquad (4)$$

(This expression was first appeared in the author's 1999 e-print paper [11] in the attempts on the explanation of the cosmological constant problem by the electroweak and Planck scale.). If we put $m_{EW}$ to about 100 GeV, the value of $\lambda$ is

$$\lambda \sim 10^{-82}[GeV]^2 = 10^{-52}[m]^{-2} \qquad (5)$$

(we use $1GeV \sim 10^{-15} m^{-1}$). If we use a dimensionless Hubble constant $h = 0.65$ (recent observational results give the Hubble constant in the range 0.58-0.72 [12] ), our calculated cosmological constant will be equivalent to about $0.7 \rho_c$ ( $\rho_c = 3H_0^2/8\pi G$ is the critical mass density when $\lambda = 0$). It is in excellent agreement with the Type Ia SN observation data.

Besides the explanation of the present $\lambda$ value, this theory has important implications on the early development of the universe. We can first put equation (4) into a more general form as

$$\lambda = k\rho_v \sim \left(\frac{m_T^7}{M^5}\right) \qquad (6)$$

where $m_T$ is the VEV mass scale of the scalar field in different transition stage in the early universe. The $\lambda$ value for the 3 transition stages, the Planck stage, GUT stage and the electroweak stage were then equal to $M^2$, $m_{GUT}^7/M^5$ and $m_{EW}^7/M^5$ respectively. If we believe that the mass energy density (including radiation and matter) of the universe is came from the change of the macroscopic vacuum energy density, we expect that at the moment just after the GUT and the electroweak transition the matter density were as $M^4$ and $m_{GUT}^7/M^3$ respectively. One of the point remains uncertain and cannot be given by this theory is the mass energy density at Planck time. If it was comparable to the vacuum energy at that time, the $\lambda$ effect would dominated the mass energy effect at about $10^{-42} s$ (estimated by Friedmann model) and the universe would be fast cooled to GUT transition temperature by the accelerated expansion due to the cosmological constant. This made the GUT transition stages came earlier than in the hot Friedmann universe (at $10^{-35}s$).

After GUT transition, the mass energy density would be equal to $M^4$ and $\mathbf{\lambda} = m_{GUT}^7/M^5$. The universe was then dominated by mass and continuous cooling by the usual Friedmann expansion. Assuming that the universe was radiation dominated and the relation between the cosmic temperature and time was as $T \propto t^{-1/2}$ (we neglect the cosmological constant effect before it dominated the expansion for simplicity), we expect that the universe would be dominated by the cosmological constant at the time around $10^{-28}s$ (This is estimated by the calculation that the mass energy density is greater than vacuum energy density by 28 order of magnitude at around $10^{-42}s$ and this corresponding to the change of temperature of about 7 order of magnitude. Therefore, the $\mathbf{\lambda}$ would be dominated at $10^{14} \times 10^{-42}s = 10^{-28}s$.). The universe would then be fast cooled by the accelerated expansion in a short time. At the time just before the electroweak transition, the $\mathbf{\lambda}$ value was then about 40 order of magnitude larger than the mass energy density and 91 order of magnitude larger than the present $\mathbf{\lambda}$ value. Such huge cosmological constant effect might cause an extreme large expansion of the universe in that period. This is what we commonly called the "inflation"! Although it is also driven by a huge $\mathbf{\lambda}$ value, the acceleration process is not due to the false vacuum as other inflation models [9] but is an intrinsic property of the space-time "condensate". As in the above estimated time of transition, we have not consider the additional acceleration effect by the $\mathbf{\lambda}$ before it become dominated. It is then reasonable to expect that the inflation also occurred at around the time order $10^{-28}s$. One can find that $\mathbf{\lambda}^{1/2}t \sim 1$ at that time. The inflation would be ended at the completion of the electroweak transition. This means that it also came earlier than the expected time of about $10^{-10}s$ in the hot Friedmann universe. As in the usual inflation model, the inflation also occurred in the period between the GUT and electroweak transition but the cosmic time was different.

After the electroweak transition, the mass energy became $m_{GUT}^7/M^3$ and $\lambda = m_{EW}^7/M^5$. The cosmological constant was dropped to the present value, the inflation then stopped and the universe became matter dominated again. When the matter density of the universe continuously decreased by the expansion of the universe, the $\lambda$ value becomes dominate again as the present cosmological observation [1,2]. If the radiation dominated universe ended at around $10^{12}s$ and enter the matter dominate stage, the dilution factor for the mass energy density due to the cosmic expansion from $10^{-28}-10^{12}s$ was about $10^{80}$. The additional factor in the matter stage up to now contribute about $10^{10}$. That means the estimated present mass energy density is $10^{-90} \times m_{GUT}^7/M^5 = 10 \times m_{EW}^7/M^5 = 10\lambda$. It is pretty close to the observation that the mass energy density about the same order as the vacuum energy density in present universe. The estimation is a bit large and may due to uncertainty in estimating the ending time of the electroweak transition. It is because only half an order of magnitude change can cause such deviation. If we believe that there is no further phase transition of the space-time "condensate" in future, our universe will then dominate by $\lambda$ value forever. Up to now, we know that the universe evolution stages starting from Big Bang might be alternatively dominated by cosmological constant and matter density in different transition stages and the cosmological constant will win the process finally. Also, from the mass energy density estimation above, it seems that $\rho_m \sim \rho_v$ in the present universe may be just a coincidence, not due to underlying physical theory. This result supports the anthropic principle [5].

From the above theory, we find that the cosmological constant problem can be resolved by postulating that space-time is discrete in nature with its phase transition properties described by the scalar field. The cosmological constant calculated by our theory is in excellent agreement with the Type Ia SN observation data. Since the $\lambda$ value is dependent on the VEV of the scalar field and therefore it is closely related to the phase transition of the space-time "condensate". The evolution of the universe including the Big Bang might be a series of phase transition of this "condensate". Our theory automatically gives out the inflation process in the early universe but the stage of acceleration is different from other models of inflation. It explains the cosmological constant problem and the inflation mechanism together in a single simple theory. The evolution of the universe is found to be alternatively dominated by the cosmological constant and the mass density at different transition stages. Our calculation shows that $r_m \sim r_v$ in the present universe. It is then just a coincidence, not due to underlying physical theory. Provided that no further phase transition will occur, our theory predicts that the universe will be dominated by the $\lambda$ value forever. One may find that all the above results is not achieved by fine tuning parameters but follows automatically from our postulates. In our theory, the scalar field also become more physical than just an unknown vacuum potential but is the phase parameter and wavefunction of the space-time "condensate". The divergence problem of quantum gravity can be automatically solved because the space-time itself is not a fundamental field but a collective effect of a more fundamental Higgs process. The above arguments shows that the space-time structure can be a more complicated structure then just a continuous mathematical space so that an evolution on the space-time concept is therefore necessary.